\RequirePackage{fix-cm}
\documentclass[twocolumn]{svjour3}          
\smartqed  
\usepackage{graphicx}
\usepackage{amsmath, amsfonts}
\usepackage{enumitem}
\usepackage{subcaption}
\usepackage{hyperref}
\def\makeheadbox{\relax}
\newtheorem{alg}{Algorithm}
\journalname{Statistics and Computing}
\begin{document}

\title{Ensemble sampler for infinite-dimensional inverse problems
}


\author{Jeremie Coullon \and
        Robert J. Webber 
}


\institute{Jeremie Coullon \at
              Lancaster University,
              LA1 4YF United Kingdom \\
              \email{jeremie.coullon@gmail.com} \\
           \and
           Robert J. Webber \at
              New York University,
              10012 New York, United States \\
              \email{rw2515@nyu.edu}
}

\date{}

\maketitle

\begin{abstract}
We introduce a new Markov chain Monte Carlo (MCMC) sampler for infinite-dimensional inverse problems.
Our new sampler is based on the affine invariant ensemble sampler,
which uses interacting walkers to adapt to the covariance structure of the target distribution.
We extend this ensemble sampler
for the first time to
infinite-dimensional function spaces,
yielding a highly efficient gradient-free MCMC algorithm.
Because our new ensemble sampler does not require gradients or 
posterior covariance estimates, 
it is simple to implement and broadly applicable.

\end{abstract}

\section{Introduction}
\label{intro}

In many Bayesian inverse problems,
Markov chain Monte Carlo (MCMC) methods are needed to approximate distributions on infinite-dimensional function spaces, for example
in groundwater flow \cite{Iglesias_2014}, medical imaging \cite{dunlop2016bayesian}, and traffic flow \cite{coullon2020mcmc}.
Yet designing efficient MCMC methods for function spaces has proved challenging.

The earliest proposed sampler for function spaces was the preconditioned Crank-Nicolson algorithm (PCN, \cite{beskos2008mcmc}).
PCN is easy to code and broadly applicable,
but it is not always efficient.
When sampling from a posterior distribution that
is poorly scaled or multimodal,
PCN can require a huge number of samples to 
accurately calculate statistics \cite{cotter2013mcmc}.

Recent
gradient-based
MCMC methods
\cite{cotter2013mcmc,cui2014likelihood,beskos2017geometric},
preconditioned MCMC methods \cite{zhou2017hybrid,rudolf2018generalization},
and SMC methods \cite{Kantas2014SMC}
have improved on the computational efficiency of PCN.
However, these new samplers require gradients or posterior covariance estimates
that may be challenging to obtain.
Calculating gradients is difficult or impossible in many 
high-dimensional inverse problems
involving a numerical integrator with a 
black-box code base
\cite{chen2016accelerated}.
Additionally, accurately estimating posterior covariances can require a lengthy pilot run or adaptation period \cite{roberts2009examples}.
These concerns raise the question:
is there a functional sampler
that outperforms PCN without requiring gradients
or posterior covariance estimates?

To address this question,
we turn to the literature on finite-dimensional MCMC.
In finite-dimensional spaces,
there is a gradient-free sampler that 
avoids explicit covariance estimation
yet adapts naturally to the covariance structure of the sampled distribution.
This sampler, called the affine invariant ensemble sampler (AIES, \cite{goodman2010ensemble}), 
is easy to tune,
easy to parallelize,
and efficient at sampling
spaces of moderate dimensionality ($d \leq 20$).
AIES is used extensively due to its implementation in the popular \texttt{emcee} package for python \cite{foreman2013emcee}.

The main contribution of this work is to propose a new functional ensemble sampler (FES)
that combines PCN and AIES.
To apply this new sampler, we first calculate
the Karhunen–Lo{\`e}ve (KL) expansion for the Bayesian prior distribution, assumed to be Gaussian and trace-class.
Then, we use AIES to sample the posterior distribution on the 
low-wavenumber KL components and use 
PCN to sample the posterior distribution on the high-wavenumber KL components.
Alternating between AIES and PCN updates,
we obtain our functional ensemble sampler 
that is efficient
and easy to use, without requiring detailed knowledge of the target distribution.

In past work,
several authors have proposed splitting the
Bayesian posterior into low-wavenumber and high-wavenumber components and then applying enhanced sampling to the low-wavenumber components \cite{law2014proposals,Kantas2014SMC,cui2016dimension,beskos2017geometric,beskos2018multilevel}.
Yet compared to these other samplers, FES is unique in its simplicity and broad applicability.
FES does not require any derivatives,
and the need for derivative-free samplers has previously been emphasized in \cite{chen2016accelerated,hu2017adaptive,zhou2017hybrid,beskos2018multilevel}.
FES also eliminates the requirement for posterior covariance estimates.
Lastly, FES is more efficient than other gradient-free samplers
in our tests.

In two numerical examples, we apply FES
to challenging inverse problems that involve 
estimating a functional parameter
and one or more scalar parameters.
In our first example, we consider the advection equation
\begin{equation}
\begin{cases}
    \frac{\partial \rho}{\partial t} + c \frac{\partial \rho}{\partial x} = 0, & t > 0, \\
    \rho = \rho_0, & t = 0.
\end{cases}
\end{equation}
We simultaneously estimate the advection speed $c$ and the initial condition $\rho_0$ from a set of noisy observations.
We compare the performance of PCN and FES
and find that PCN mixes slowly
because $c$ and $\rho_0$ are highly correlated under the posterior distribution.
In comparison,
FES mixes more quickly,
reducing integrated autocorrelation times \cite{sokal1997monte} by two orders of magnitude.

In our second example, we consider the Langevin diffusion
\begin{equation}
\begin{cases}
    \mathop{dX} = P \mathop{dt}, & t > 0, \\
    \mathop{dP} = -\alpha X \mathop{dt} + \sigma \mathop{dW}, & t > 0, \\
    X = P = 0, & t = 0.
\end{cases}
\end{equation}
We simultaneously estimate the drift parameter $\alpha$, the diffusion parameter $\sigma$,
and the posterior path $\left(X_t\right)_{0 \leq t \leq 10}$
from noisy observations.
We compare the performance of PCN, FES, and an alternative derivative-free sampler \cite{zhou2017hybrid}
that explicitly estimates the posterior covariance matrix.
We conclude that FES is the fastest available gradient-free sampler
for this challenging, multimodal test problem.

The rest of the paper is organized as follows.
Section \ref{sec:background} reviews the PCN and AIES samplers,
Section \ref{sec:ensemble} introduces the new ensemble sampler for function spaces,
Section \ref{sec:examples} presents numerical examples,
and Section \ref{sec:conclusion} concludes.
Code to reproduce the examples is available on Github\footnote{\url{https://github.com/jeremiecoullon/functional_ensemble_sampler}}.

\section{Background on MCMC samplers}{\label{sec:background}}

In this section, we explain 
why it is difficult to 
approximate a Bayesian posterior distribution 
on an infinite-dimensional function space.
Then, we describe the
preconditioned Crank-Nicolson sampler
(PCN, \cite{beskos2008mcmc})
which can be used for this approximation task.
Lastly, we describe the affine invariant ensemble sampler (AIES, \cite{goodman2010ensemble}),
an efficient sampler for finite-dimensional spaces
that has not previously been extended to the infinite-dimensional setting.

\subsection{Infinite-dimensional inverse problems}

In a typical infinite-dimensional Bayesian inverse problem,
the goal is estimating a posterior distribution
\begin{equation}
\label{eq:standard}
\pi(\mathop{du}) \propto \exp\left(\phi(u)\right) \pi_0(\mathop{du}),
\end{equation}
where $u$ is a square-integrable function on a domain 
$\Omega \subseteq \mathbb{R}^d$,
$\phi(u)$ is a log-likelihood functional, 
and $\pi_0 = \mathcal{N}(0, C)$ is a Gaussian prior distribution.

To estimate $\pi$,
we must select a finite-dimensional approximation space
and then sample $\pi$ restricted to 
this space.
However,
ensuring high accuracy with this approach
is difficult.
To accurately calculate
statistics of the posterior distribution,
a high-dimensional approximation space 
is needed.
Yet, as we increase the dimensionality,
the acceptance probability for a standard MCMC sampler,
such as the Metropolis random walk sampler \cite{metropolis1953equation},
sinks to zero.
Hence, the MCMC sampler takes an increasingly long time to move anywhere,
and sampling from $\pi$ becomes tediously slow \cite{cotter2013mcmc}.

\subsection{Preconditioned Crank-Nicolson \label{sub:PCN}}

PCN solves the problem of vanishing acceptance probabilities
by proposing MCMC moves that are \emph{always} accepted under the Gaussian prior distribution.
Because of this stability property,
even as we increase the dimensionality of the approximation space,
the average acceptance probability remains bounded away from zero \cite{hairer2014spectral}.

Starting from a position $U$, 
the PCN update is
\begin{equation}
    \tilde{U} = \sqrt{1-\omega^2} U + \omega \xi,
\end{equation}
where $\xi \sim \mathcal{N}\left(0, C\right)$ is a random draw from the Gaussian prior
and $\omega \in \left(0, 1\right]$ is a step size parameter. 
If $\omega \ll 1$, the proposal is a small perturbation of the position $U$, 
whereas if $\omega = 1$ the proposal is independent from $U$.
The main computational cost of PCN then comes from evaluating the acceptance probability
\begin{equation}
\min\left\{1, \exp\left(\phi(\tilde{U}) - \phi(U)\right)\right\},
\end{equation}
which requires calculating the log-likelihood functional at the proposed parameter value $\tilde{U}$.

PCN is a simple, widely applicable
approach
that requires little more than making inexpensive proposals
and evaluating the log-likelihood at the proposed parameter values.
However, the main limitation of PCN is the slow convergence of statistics
when the posterior distribution is poorly scaled or multimodal.
This slow convergence has led to myriad
efforts to improve on PCN's sampling speed \cite{cotter2013mcmc,cui2014likelihood,Kantas2014SMC,beskos2017geometric,zhou2017hybrid,rudolf2018generalization}, but the available methods require gradients or covariance estimates that can be difficult to obtain.

\subsection{Affine invariant ensemble sampler}

The affine invariant ensemble sampler (AIES, \cite{goodman2010ensemble}) 
is a finite-dimensional MCMC sampler with the remarkable property of \emph{affine invariance}.
Affine invariance means that the sampler remains completely unchanged
if the state space is stretched, compressed,
or translated by an affine transformation $x \mapsto Ax + b$.
Because of this property,
AIES efficiently samples from
distributions that are wide in some directions and narrow in other directions.
These ``poorly scaled" distributions
would cause problems for other samplers,
but they do not compromise the performance of AIES.

To sample from a density $\pi$ on $\mathbb{R}^M$,
AIES generates an ensemble of walkers
$\overrightarrow{X} = (X_1, ... X_L)$
that is invariant with respect to
the product density $\pi\left(x_1\right) \cdots \pi\left(x_L\right)$ on $\mathbb{R}^{ML}$.
To update the ensemble, 
AIES proposes sliding one walker
toward or away from another walker.
Then, AIES accepts or rejects the proposal
according to a Metropolis-Hastings step.
The proposals and acceptance probabilities are invariant under affine transformations,
so the scheme is affine invariant overall.

To perform the AIES proposal step,
we randomly choose a walker $X_i$ 
and a second walker $X_j \neq X_i$.
Then, we propose moving the walker $X_i$
to the new position
\begin{equation}
\label{eq:proposal}
    \tilde{X}_i = X_i + \left(1 - Z\right) \left(X_j - X_i\right),
\end{equation}
where $Z$ 
is a random number in an interval $\left[1 \slash a, a\right]$, chosen with density $g\left(Z\right) \propto 1 \slash \sqrt{z}$.
Typically, $a = 2$ in applications, but more generally $a$ is a parameter that modulates the step size.
The main computational cost of AIES then comes from evaluating the acceptance probability
\begin{equation}
\min \left\{ 1, Z^{M-1} \frac{\pi(\tilde{X}_i)}{\pi(X_i)}  \right\},
\end{equation}
which requires calculating 
the density $\pi$ at the proposed position $\tilde{X}_i$.

AIES is a popular and efficient sampler 
for low- and moderate-dimensional densities
($M \leq 20$).
AIES would not typically be an efficient sampler for higher-dimensional denities.
However, in the sections to follow, we explain how AIES can be applied to a low-dimensional subspace of an infinite-dimensional function space, thereby improving the sampling compared to PCN.

\begin{remark}{\label{rem:parallel}}
A parallel implementation of AIES is available in the \texttt{emcee} package for python \cite{foreman2013emcee}. 
In this version of AIES, we split the walkers into two groups
and sample in two stages.
Initially, we select walkers from the first group
and slide these walkers toward or away
from walkers in the second group.
Then, we select walkers from the second group
and slide these walkers toward or away from walkers in the first group.
By splitting the walkers into two groups,
we can conduct AIES in parallel across multiple processors,
helping to spread out the computational cost.
\end{remark}

\section{New ensemble sampler}
{\label{sec:ensemble}}

In this section, we describe the Karhunen-Lo{\`e}ve (KL) expansion, which is helpful tool for 
constructing functional MCMC samplers.
Then, we introduce our new functional ensemble sampler (FES) and 
discuss its main properties.

\subsection{KL expansion}

The KL expansion \cite{stuart2010inverse} is a 
rapidly converging basis expansion for a random function $\xi$ drawn
from a trace-class Gaussian distribution 
$\mathcal{N}\left(0, C\right)$.
The KL expansion decomposes $\xi$ into a linear combination of ``KL modes" $\eta_1, \eta_2, \ldots$, 
which are eigenfunctions of the covariance operator $C$.
Thus, the KL expansion takes the form 
\begin{equation}
\label{eq:kl}
    \xi = \sum_{i=1}^{\infty} \left<\eta_i, \xi\right> \eta_i,
\end{equation}
where $\left<\cdot, \cdot\right>$ denotes the inner product in $L^2\left(\Omega\right)$.
Because $\xi$ is a Gaussian with mean zero,
the KL components $\left<\eta_i, \xi\right>$ are 
independent Gaussians with mean zero
and variances $\lambda_1 \geq \lambda_2 \geq \cdots$
that are determined by the eigenvalues of $C$.

The KL expansion converges as rapidly as possible in the sense
of minimizing the mean squared error
\begin{equation}
    \mathbb{E}_{\xi \sim \mathcal{N}\left(0, C\right)} \left\lVert \xi - \sum_{i=1}^L \left<\eta_i, \xi\right> \eta_i \right\rVert^2_{L^2\left(\Omega\right)},
\end{equation}
for any truncation threshold $L \geq 1$.
Because the KL expansion converges so rapidly,
the low-wavenumber modes explain most of the variance in $\xi$.
For example, if $\xi$ is a Brownian motion on $\left[0, 1\right]$,
the eigenfunctions of the covariance operator are
\begin{equation}
    \eta_i\left(t\right) = \sqrt{2} \sin\left(\left(i - \frac{1}{2}\right) \pi t\right), \quad i = 1, 2, \ldots,
\end{equation}
and the  eigenvalues
are $\lambda_i = \left(i- \frac  {1}{2}\right)^{-2} \pi^{-2}$.
Hence, the five lowest-wavenumber KL
modes account for $96\%$ of the variance in $\xi$,
while the high-wavenumber modes account for just $4\%$ of the variance.

We now consider the implications of the KL expansion
for Bayesian inference.
In a Bayesian inverse problem with a Gaussian prior,
we can decompose a functional parameter $U$ in terms of the KL modes
\begin{equation}
    U = \sum_{i=1}^{\infty} U_i \eta_i, \quad U_i = \left<\eta_i, U\right>.
\end{equation}
Under the prior distribution $\pi_0 = \mathcal{N}\left(0, C\right)$,
the $U_i$ components are independent Gaussians.
Under the posterior distribution $\pi\left(\mathop{du}\right) \propto \exp\left(\phi\left(u\right)\right) \pi_0\left(\mathop{du}\right)$,
the $U_i$ components have an unknown distribution
that must be approximated through sampling.

Although the $U_i$ components
have an unknown posterior distribution,
the prior distribution restricts the values these variables
can take.
The high-wavenumber components are narrowly peaked Gaussians 
under the prior,
so they are constrained to be nearly Gaussian with a low variance under the posterior.
In contrast, the low-wavenumber components
are less constrained,
so
the posterior distribution on these components can become stretched, pinched, or otherwise distorted by the likelihood function.

The KL coordinates divide an infinite-dimensional 
inverse problem into a simple sampling part and a challenging sampling part.
Sampling the high-wavenumber components is comparatively simple.
The prior and posterior distributions 
on these components are nearly the same,
enabling PCN to sample efficiently.
In contrast, 
sampling the low-wavenumber components
is more challenging.
The posterior distribution on these components
may be poorly scaled or multimodal,
causing difficulties for PCN.

\subsection{Functional ensemble sampler (FES)}

To efficiently sample from function spaces,
we propose a
Metropolis-within-Gibbs sampler
that uses
AIES on the low-wavenumber KL components
and PCN on the high-wavenumber KL components.
We call this algorithm the functional ensemble sampler (FES) and provide pseudocode for the method below.

\begin{alg}[Functional ensemble sampler]{\label{alg:fes_algorithm}}
~
\newline
To sample a distribution $\pi(\mathop{du}) \propto \exp\left(\phi(u)\right) \pi_0(\mathop{du})$
where
$\pi_0 = \mathcal{N}\left(0, C\right)$, perform the following steps:
\begin{enumerate}[leftmargin = *]
    \item Identify a matrix $J$ whose columns are the first $M$ eigenvectors of $C$.
    Set $P = J J^T$
    and $Q = I - J J^T$.
    \item Initialize an ensemble of walkers $\left(X_1^0, ... X_L^0\right)$.
    \item For $\tau = 0, 1, \ldots$:
    \begin{enumerate}[leftmargin = .5cm]
        \item For $i = 1, \ldots, L$: 
        \begin{enumerate}[leftmargin = .5cm]
        \item Randomly choose a walker $X_j^{2\tau} \neq X_i^{2\tau}$.
        \item Propose the update 
        \begin{equation}\label{eq:FES_alg_AIES_proposal}
            \tilde{X}_i^{2\tau} = X_i^{2\tau} + \left(1 - Z\right) P \left(X_j^{2\tau} - X_i^{2\tau}\right),
        \end{equation}
        where $Z \in \left[1\slash a, a\right]$ has density $g\left(z\right) \propto 1 \slash \sqrt{z}$.
        \item
        Set $X_i^{2\tau} = \tilde{X}_i^{2\tau}$ with probability 
        \begin{equation}
            \min\left\{1, Z^{M-1} \frac{\pi\left(\tilde{X}_i^{2\tau}\right)}{\pi\left(X_i^{2\tau}\right)}\right\}.
        \end{equation}
        \end{enumerate}
        \item Set
        $\left(X_0^{2\tau + 1}, \ldots, X_L^{2\tau + 1}\right) = \left(X_0^{2\tau}, \ldots, X_L^{2\tau}\right)$.
        \item For $i = 1, \ldots, L$:
        \begin{enumerate}[leftmargin = .5cm]
        \item Propose the update
        \begin{equation}\label{eq:FES_alg_PCN_proposal}
            \tilde{X}_i^{2\tau + 1} = P X_i^{2\tau + 1} + Q \left(\sqrt{1 - \omega^2} X_i^{2\tau + 1} + \omega \xi\right),
        \end{equation}
        where $\xi \sim \mathcal{N}\left(0, C\right)$.
        \item Set $X_i^{2\tau+1} = \tilde{X}_i^{2\tau+1}$ with probability
        \begin{equation}
            \min\left\{1, \exp\left(\phi\left(\tilde{X}_i\right) - \phi\left(X_i\right)\right)\right\}.
        \end{equation}
        \end{enumerate}
        \item Set
        $\left(X_0^{2\tau + 2}, \ldots, X_L^{2\tau + 2}\right) = \left(X_0^{2\tau + 1}, \ldots, X_L^{2\tau + 1}\right)$.
    \end{enumerate}
\end{enumerate}
\end{alg}

\subsection{Properties of FES}\label{sec:limitationas_extensions_comparisons}

FES is a novel method for
function space sampling,
which enhances the standard PCN approach.
FES remains stable as we refine the functional discretization, similarly to PCN.
However, compared to PCN,
we can tune FES to achieve faster mixing.

The main tuning parameter in FES is $M$,
which controls how many KL coordinates 
are included in the AIES sampling.
When $M = 0$, no AIES sampling is performed, 
so the algorithm reduces to PCN.
As $M$ increases, FES begins to outperform PCN.
However, if $M$ increases past $20$, the performance
deteriorates, since AIES is only an efficient sampler for subspaces of dimension $20$ and lower.

The precise number of KL coordinates
to include in the AIES sampling is a tuning decision,
with the optimal number depending on the estimation problem.
However, based on our numerical tests,
we recommend setting $M = 5$ as a default
and then adjusting $M$ during the early stages of the sampling to be as small as possible while ensuring the PCN sampler can take large steps ($\omega \geq 0.5$)
with acceptance rate $\geq 20\%$.



To explain the limitations of FES,
we recall the idea of a \emph{likelihood-informed subspace} (LIS), originally introduced in \cite{cui2014likelihood}.
An LIS is a low-dimensional linear subspace 
in which prior and posterior marginal distributions differ substantially.
Moreover, conditioning on 
an LIS ensures that differences between prior and posterior distributions become small.
An LIS is useful for constructing efficient MCMC algorithms,
because enhanced sampling is needed on the LIS
but PCN provides efficient updates
in directions orthogonal to the LIS
\cite{cui2014likelihood,cui2016dimension,beskos2018multilevel}.

FES relies on the assumption that 
the $20$ 
lowest-wavenumber KL components
contain an LIS.
This assumption is 
often but not always satisfied in practice.
By considering a sufficiently large number of KL coordinates,
we can always find an LIS.
However, the required number of coordinates
may be larger than $20$.
For example, a large number of KL coordinates is needed if the
posterior distribution emphasizes solutions that are not very smooth,
which can happen
if the observational noise in the problem
is small. 
If the required number of KL coordinates is higher than $20$, 
FES may no longer provide an efficient sampling solution,
although it is still not slower than PCN.

Ideally, we would extend FES by applying
AIES directly to a likelihood-informed subspace
and applying PCN to the complementary subspace.
However, to our knowledge, all the available methods for identifying an LIS
require calculating gradients \cite{cui2016dimension} or posterior covariance matrices \cite{beskos2018multilevel}.
Developing broadly applicable tools for identifying an LIS remains an
issue for future research.

Other extensions to FES are also possible.
Whereas Algorithm \ref{alg:fes_algorithm} presents a sequential implementation of FES,
there is also a parallel implementation
using the modified AIES sampling discussed in Remark \ref{rem:parallel}.
Another extension to FES involves jointly sampling
functional and scalar parameters in a Bayesian inverse problem.
Indeed, it is straightforward to include 
additional scalar parameters in the AIES subspace,
as we demonstrate through numerical examples in Section \ref{sec:examples}.
We regard this extension of FES as especially useful, since there is often simultaneous uncertainty around functional and scalar parameters in a model.

Lastly, we compare FES to
the ``hybrid sampler'' of Zhou and coauthors \cite{zhou2017hybrid}.
The hybrid sampler is a gradient-free method
that uses PCN to sample the high-wavenumber KL components
and uses Gaussian random walk proposals to sample the low-wavenumber KL components.
In the hybrid sampler, the covariance of the Gaussian perturbations
is adaptively tuned based on the estimated posterior covariance matrix.

We find in our experiments that the hybrid sampler can be very efficient when the posterior distribution is nearly Gaussian and the posterior covariance matrix is accurately estimated 
(even slightly more efficient than FES).
However, limitations of the hybrid sampler include
sensitivity to non-Gaussian posterior distributions
and long adaptation periods needed to achieve peak performance.
As we show in 
Section \ref{sec:examples},
FES addresses both of these limitations.
FES is a fast sampler
for many non-Gaussian distributions, 
and FES is efficient over short sampling runs.

\section{Numerical examples}{\label{sec:examples}}

In this section, we apply FES 
to two challenging inverse problems
involving functional and scalar parameters.
For both problems, we fix the AIES step size to $a = 2$, as recommended in \cite{foreman2013emcee}, and we tune the PCN step size $\omega$ to give an acceptance rate of $20\%$.
We remove the first $10\%$ of each trajectory as burn-in, and we run the trajectories 
at least $100$ times as long as the integrated autocorrelation time
to ensure robust statistics
\cite{sokal1997monte}. 

\subsection{Advection equation}

We first consider the advection equation $\frac{\partial\rho}{\partial t} + c\frac{\partial \rho}{\partial x} = 0$,
a simple first-order PDE that is a prototype
for more general hyperbolic PDEs.
Given an initial condition $\rho_0(x)$ and a wave speed $c \in \mathbb{R}$,
the solution to the advection equation
can be written explicitly as
\begin{equation}\label{eq:equation_advection}
    \rho(x, t) = \rho_0(x-ct).
\end{equation}

We aim to 
recover the initial condition and wave speed from noisy observations of flow.
Flow is the product of density and velocity, given by the equation $q = \rho c$.
When flow is the only quantity observed,
the initial condition and wave speed
become highly correlated in the posterior,
making the MCMC sampling difficult.


In our Bayesian model, we set a 
$\mathrm{Unif}\left(0, 1.4\right)$ prior on $c$
and a Gaussian prior on $\rho_0$ with mean $100$ and covariance function 
\begin{equation}
k(x, x') = 130 \exp\left(-\frac{1}{2} \left(x-x^{\prime}\right)^2\right).
\end{equation}
We generate a true solution to the PDE by setting $c_{\mathrm{true}} = 0.5$
and drawing $\rho_0$ according
to the Gaussian prior.
Then, we generate observations 
of the flow at locations
$x = 2$, $6$, and $10$
and times $t = 1$, $1.5$, and $2$,
subject to independent
$\mathcal{N}\left(0, 0.04\right)$
observational noise.

To approximate the posterior distribution on $\rho_0$ and $c$,
we apply FES using $L = 100$ walkers.
During the initialization,
we independently sample the walkers from a ball around the posterior mode, as recommended in \cite{foreman2013emcee}.
We discretize $\rho_0$
using $200$ grid points,
equally spaced between $x = 0$ and $x = 10$.

In FES applications, we recommend choosing the AIES subspace 
to be as low-dimensional as possible,
while ensuring that PCN can take large steps ($\omega \geq .5$)
with a high acceptance rate ($\geq 20\%$).
Here, we empirically support this recommendation
by evaluating the performance of FES when
the AIES subspace includes the wave speed parameter $c$
as well as $M=0$, $1$, $5$, $10$, or $20$ of the lowest-wavenumber KL components.

As our first conclusion from this comparison, 
we find that we can take larger PCN steps
with a $20\%$ acceptance rate
if we choose $M$ to be large.
We report the precise PCN step sizes in Table \ref{table:advection_omega_table},
which reveals that a PCN step size
$\omega \geq .5$ is possible
whenever $M \geq 10$.

\begin{table}[h!]
\centering
\begin{tabular}{ |l|l|l|l|l|l| }
 \hline
 \multicolumn{6}{|c|}{PCN step size} \\
 \hline
  & $M{=}0$ & $M{=}1$ & $M{=}5$ & $M{=}10$ & $M{=}20$  \\
 \hline
 \(\omega\)   & 0.04 & 0.05 & 0.15 & 0.60 &  1.00 \\
 \hline
 
\end{tabular}
\caption{\label{table:advection_omega_table} PCN step size for various FES trials.}
\end{table}

\begin{figure}[h!]
\centering
\includegraphics[clip, scale=0.32]{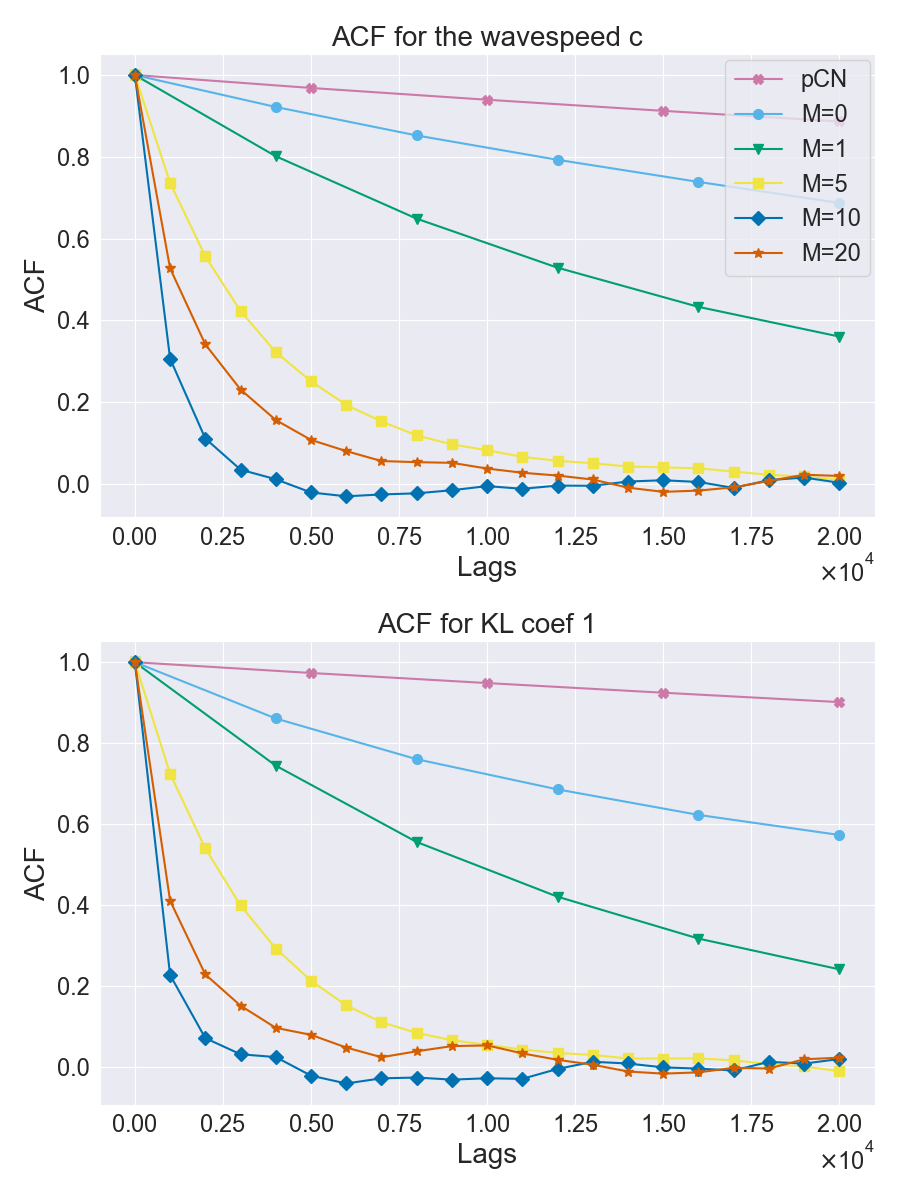}
\caption{ACF curves for wave speed $c$ and 
the first KL coefficient.}
\label{advection_ACF}
\end{figure}

As our second conclusion, we verify that choosing $M = 10$ 
leads to the most efficient sampling.
We report the autocorrelation functions (ACFs)
and integrated autocorrelation times (IATs)
for various observables
in Figure \ref{advection_ACF} and Table \ref{table:advection_IAT_table}.
For comparison purposes, we also report the ACFs and IATs
from a standard PCN-based sampler.
With the optimal parameter $M = 10$,
we find that FES reduces the IATs by two orders of magnitude compared to PCN.

\begin{table}[h!]
\centering
\begin{tabular}{ |l||l|l|l|l|l|l|l|  }
 \hline
 \multicolumn{7}{|c|}{Integrated autocorrelation time $\div\, 1000$} \\
 \hline
  & PCN & $M{=}0$ & $M{=}1$ & $M{=}5$ & $M{=}10$ & $M{=}20$  \\
 \hline
 $c$ 
 & $360$ 
 & $130$ 
 & $50$ 
 & $7.7$ 
 & $\boldsymbol{1.5}$ 
 & $4.3$ \\
 $\eta_1$ 
 & $390$  
 & $110$ 
 & $30$ 
 & $6.8$
 & $\boldsymbol{1.4}$
 & $3.1$ \\
 $\eta_5$ 
 & $290$ 
 & $46$ 
 & $55$ 
 & $11$ 
 & $\boldsymbol{1.1}$ 
 & $1.9$ \\
 $\eta_{15}$
 & $280$ 
 & $30$
 & $16$ 
 & $12$ 
 & $\boldsymbol{1.0}$
 & $1.2$ \\
 $\eta_{100}$ 
 & $310$ 
 & $43$ 
 & $20$ 
 & $11$ 
 & $\boldsymbol{1.1}$ 
 & $1.4$ \\
 \hline
\end{tabular}
\caption{\label{table:advection_IAT_table} 
IATs for wave speed $c$ and 
several KL coefficients
with the fastest IATs in bold.
All IATs have been divided by $1000$ to improve readability.}
\end{table}

To check that FES remains stable with increasing dimension,
we also run our experiments with a discretization
into twice as many grid points.
The IATs remain statistically indistinguishable
from those
reported in Table \ref{table:advection_IAT_table}
with relative differences of $\leq 10\%$.

\begin{figure}[h!]
\centering
\includegraphics[clip, scale=0.32]{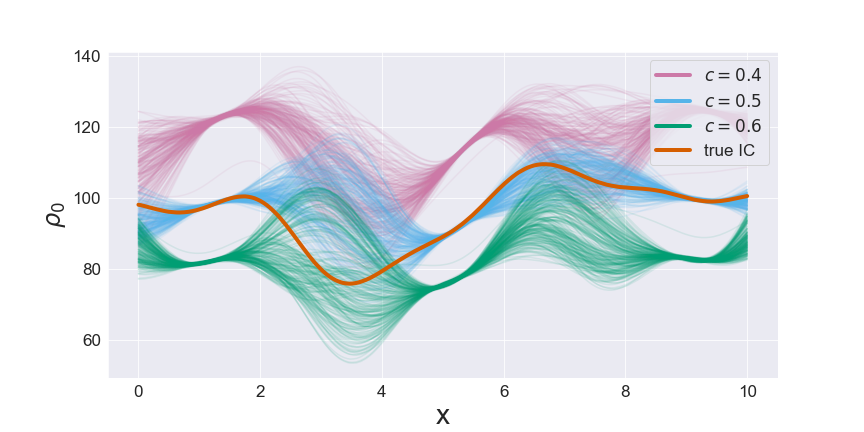}
\caption{Samples 
of $\rho_0$ conditioned on
three values of the wave speed $c$.}
\label{advection_conditional_sampling}
\end{figure}

Lastly, to help explain \emph{why} FES performs so much better than PCN,
we present Figure \ref{advection_conditional_sampling},
which shows posterior samples of $\rho_0$
conditioned on several values of $c$.
This figure reveals the strong correlation between the wave speed $c$ and the low-wavenumber components of $\rho_0$.
Since the PCN sampler does not account for this correlation structure,
large PCN updates are highly unlikely to be accepted.
In contrast, FES naturally adapts to this 
correlation structure,
eliminating the major bottleneck in the sampling.

\subsection{Path reconstruction for Langevin dynamics}{\label{sub:ex2}}

We consider 
a dynamical system
in which position $X_t$ and momentum $P_t$
evolve according to the following Langevin SDE.
\begin{equation}
\label{eq:sde}
\begin{cases}
    \mathop{dX} = P \mathop{dt}, & t > 0, \\
    \mathop{dP} = -\alpha X \mathop{dt} + \sigma \mathop{dW}, & t > 0, \\
    X = P = 0, & t = 0.
\end{cases}
\end{equation}
We aim to recover the parameters $\alpha > 0$ 
and $\sigma > 0$, 
as well as the complete path $\left(X_t\right)_{0 \leq t \leq 10}$,
based on noisy observations of position
at a few specific times.

We use the following Bayesian priors.
\begin{equation}
    \alpha \sim \text{Exp}(12),
    \quad \sigma \sim \text{Exp}(4),
    \quad W \sim \text{BM}([0, 10]).
\end{equation}
As an example of mild model misfit,
we set $X_t = \sin(4t)$ and then add $\mathcal{N}\left(0, 0.09 \right)$
observational noise at times $t = 1, 3, 5, 7$, and $9$.

To approximate the posterior path distribution,
we first infer the driving Brownian motion $\left(W_t\right)_{0 \leq t \leq 10}$ 
and the scalar parameters $\log \alpha$ and $\log \sigma$.
Then, we recover the posterior path by integrating forward the SDE \eqref{eq:sde}
using a standard Euler solver.
To discretize $X_t$ and $W_t$, 
we use $200$ equally spaced times between $t = 0$ and $t = 10$.

\begin{figure}[h!]
\centering
\includegraphics[clip, scale=0.32]{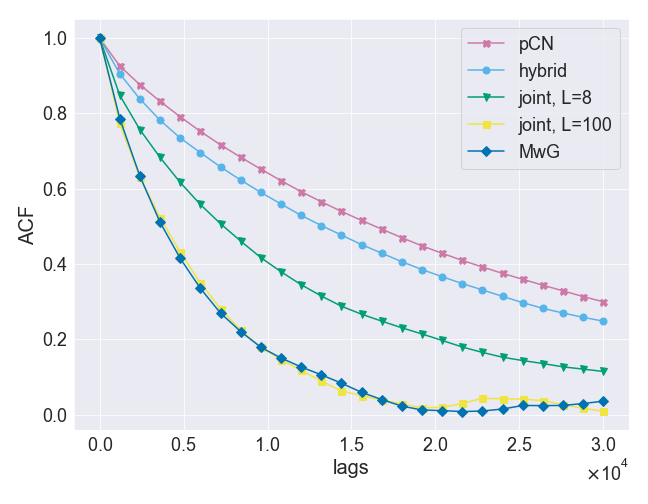}
\caption{ACF curves for the $\alpha$ parameter}
\label{fig:langevin_ACF_plots}
\end{figure}

We compare the performance of five different MCMC samplers:
\begin{enumerate}
\item A PCN-based sampler that
simultaneously proposes PCN updates for $W$
and Gaussian random walk updates for $\left(\log \alpha, \log \sigma\right)$. 
\item The ``hybrid sampler" of \cite{zhou2017hybrid}, which explicitly estimates the posterior covariance matrix.
\item A modified FES sampler with $L = 8$ walkers
and joint proposals that combine AIES and PCN moves to update all the parameters at once.
\item A modified FES sampler with $L = 100$ walkers
and joint proposals.
\item A standard FES sampler with $L = 100$ walkers.
\end{enumerate}
We initialize our samplers by drawing randomly from the Bayesian prior distribution.
After a short pilot run,
we find that $M = 5$ is a near-optimal truncation parameter, 
and we fix this parameter for all the samplers (besides PCN).
We report the IATs for the samplers in 
Table \ref{table:iat2},
and we show ACF curves in Figure \ref{fig:langevin_ACF_plots}.

\begin{table}[h!]
\centering
\begin{tabular}{ |l||l|l|p{1.2cm}|p{1.2cm}|l| }
 \hline
 \multicolumn{6}{|c|}{Integrated autocorrelation times $\div \, 1000$} \\
 \hline
  & PCN & Hybrid & Joint, $L{=}8$ & Joint, $L{=}100$ & $L{=}100$\\
 \hline
 $\log \alpha$   
 & $51$  
 & $38$ 
 & $23$ 
 & $\boldsymbol{11}$ 
 & $12$ \\
 $\log \sigma$   
 & $26$ 
 & $22$ 
 & $18$ 
 & $\boldsymbol{6.6}$ 
 & $8.1$ \\
 $\eta_1$ 
 & $5.7$  
 & $1.5$ 
 & $6.3$ 
 & $\boldsymbol{1.0}$ 
 & $1.6$ \\
 \(\eta_{10}\) 
 & $5.9$ 
 & $1.6$ 
 & $2.8$ 
 & $2.0$ 
 & $\boldsymbol{0.39}$ \\
 \(\eta_{100}\) 
 & $5.8$ 
 & $1.2$ 
 & $2.5$ 
 & $1.8$ 
 & $\boldsymbol{0.30}$ \\
 \hline
\end{tabular}
\caption{\label{table:iat2} IATs for $\log \alpha$, $\log \sigma$, and several KL coefficients with the fastest IATs in bold.
All IATs have been divided by $1000$ to improve readibility.}
\end{table}

As a first comparison, we find that FES
mixes more quickly with $L = 100$ walkers than with
$L = 8$ walkers.
$L = 8$ is the minimal possible number of walkers to ensure the AIES sampler does not get stuck in a low-dimensional subspace.
However, it is recommended to use more walkers whenever possible.
Foreman-Mackey and coauthors recommend using hundreds of walkers \cite{foreman2013emcee},
and in some applications up to 2000 walkers have been used \cite{Akeret_CosmoHammer}.

As a second comparison, we find that joint updates lead to slightly faster sampling within the AIES subspace but slower sampling in the complementary subspace, 
compared to standard FES updates.
Thus, the advantages of joint updates
versus standard Metropolis-within-Gibbs updates depend on the particular statistics being estimated.

\begin{figure}[h!]
\centering
\includegraphics[clip, scale=0.32]{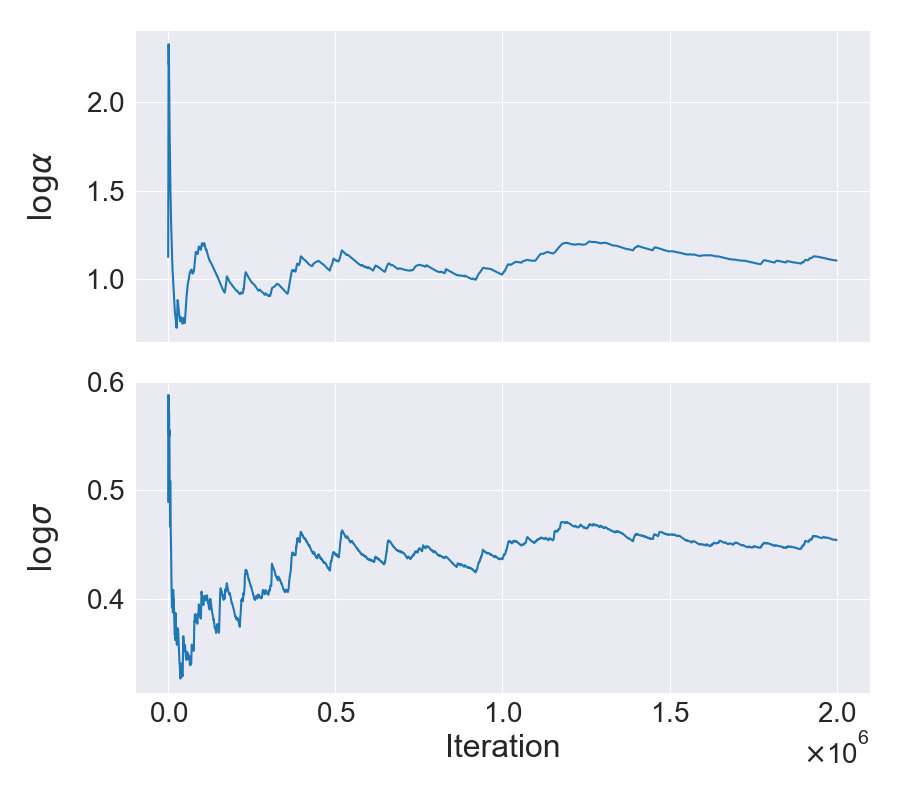}
\caption{Variance estimates for $\log \alpha$ and $\log \sigma$ using the adaptive hybrid sampler.}
\label{fig:langevin_hybrid_adaptive_plot}
\end{figure}

As a last comparison, we find that the hybrid sampler of \cite{zhou2017hybrid}
has two shortcomings that can be addressed by using FES.
First, the hybrid sampler is very slow to 
estimate the posterior covariance matrix.
Figure \ref{fig:langevin_hybrid_adaptive_plot}
reveals that more than a million iterations are needed for the estimated variances for $\log \alpha$ and $\log \sigma$ parameters to stabilize.
Since the hybrid sampler tunes its proposals based on the estimated posterior covariance matrix,
the method requires over a million iterations to achieve its peak efficiency.
In contrast, FES does not require such an adaptation period:
the dynamics remain stable from the very first iteration onwards.

\begin{figure}[h!]
    \centering
    \includegraphics[clip, scale = .35]{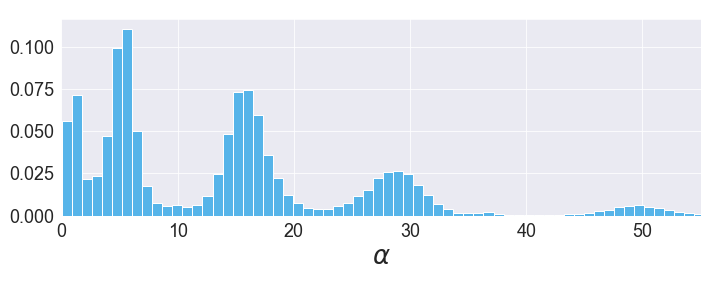}
    \caption{Posterior pdf for $\alpha$ exhibiting multimodality.}
    \label{fig:multimodal}
\end{figure}

\begin{figure}[h!]
    \centering
    \includegraphics[clip, scale = .32]{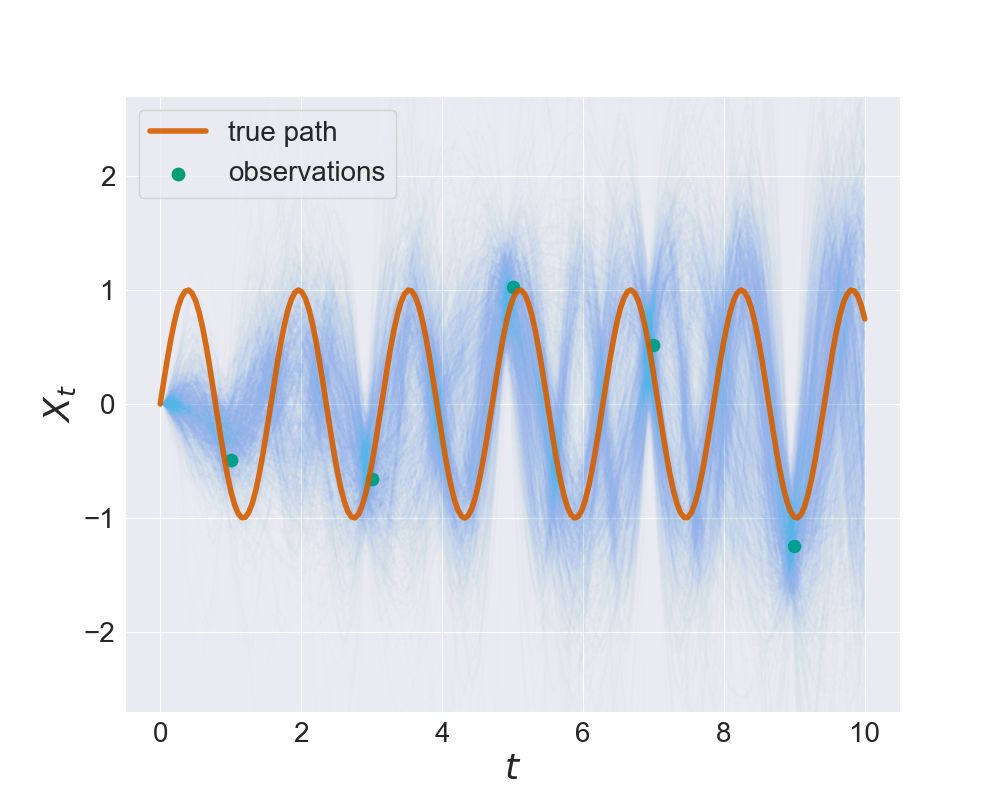}
    \caption{Posterior paths exhibiting multimodality.}
    \label{fig:paths}
\end{figure}

Second, even after the hybrid sampler has adapted to the posterior covariance structure,
mixing times for all observable are still comparatively slow.
A major obstacle limiting the efficiency of the hybrid sampler is the
\emph{multimodality} of the
posterior distribution, which is highlighted in Figures \ref{fig:multimodal} and \ref{fig:paths}.
It is very challenging for a Gaussian random walk to efficiently traverse a multimodal
distribution.
In contrast, we find in this example that FES significantly outperforms the hybrid sampler,
suggesting a robustness to multimodality that is highly desirable in applications.

\section{Conclusion}{\label{sec:conclusion}}

In this work, we introduced the functional ensemble sampler (FES).
FES requires no gradients, it is easy to code, and it is parallelizable.
These factors make FES a widely applicable sampler for infinite-dimensional
inverse problems.

In two numerical examples, we demonstrated the benefits of using FES.
First, when parameters in the posterior distribution are highly correlated,
we showed how FES can reduce integrated autocorrelation times by two orders of magnitude compared to PCN.
Second, when the posterior distribution is mildly multimodal, we showed how FES outperforms PCN and the alternative gradient-free sampler of \cite{zhou2017hybrid}.

We acknowledge two opportunities to improve the performance of FES even further.
First, 
FES sampling could be streamlined by identifying a likelihood-informed subspace
where enhanced sampling is most
essential.
Second, after isolating a low-dimensional subspace for enhanced sampling, 
we find that FES is typically an efficient sampler,
except in cases of extreme multimodality
in which FES deteriorates in its performance \cite{goodman2010ensemble} and further sampling modifications may be needed.

In conclusion, FES pushes the limits of 
the MCMC approach to solving infinite-dimensional inverse problems.
Despite having a few limitations,
the method offers a practical and powerful solution
for many sampling problems where PCN falls short,
and we recommend FES as a general-purpose gradient-free sampler.


\begin{acknowledgements}
We thank Christopher Nemeth, Gideon Simpson, and Jonathan Weare for offering useful critiques. 
JC was supported by EPSRC grant EP/S00159X/1. 
RJW was supported by the National Science Foundation award DMS-1646339
and by New York University's Dean's Dissertation Fellowship.
The High End Computing facility at Lancaster University provided computing resources.
\end{acknowledgements}

%
%

\bibliographystyle{spmpsci}      
\bibliography{bibliography}   


\end{document}